\documentclass[11pt]{article}
\usepackage[a4paper,left=2.85cm,right=2.85cm,top=2.5cm,bottom=2.5cm]{geometry}
\usepackage[colorlinks,allcolors=blue]{hyperref}
\newcommand{\articletype}[1]{}
\newcommand{\orcid}[1]{}
\providecommand{\keywords}[1]{\par\medskip\noindent\textbf{Keywords:} #1\par}
\providecommand{\roles}[1]{\section*{Author contributions}#1}
\providecommand{\ack}[1]{\section*{Acknowledgments}#1}
\providecommand{\funding}[1]{\section*{Funding}#1}
\providecommand{\data}[1]{\section*{Data availability}#1}
\usepackage{amsmath,amssymb}
\usepackage{lmodern}
\usepackage{graphicx}
\usepackage{bm}
\usepackage{booktabs}
\usepackage[numbers,sort&compress]{natbib}
\usepackage{microtype}
\usepackage{url}

\makeatletter
\let\tableinput\@@input
\makeatother
\begin{document}

\title{Design of a combined polarimetric and velocimetric measurement for viscoelastic stress, and its constitutive resolving power}
\author{Zijian Liu$^{1,*}$, Julian Olszewski$^{1,*}$, Bruce I. Gaynes$^{2}$ and Jie Xu$^{1,\dagger}$}
\date{}
\maketitle

\begin{center}\small
$^{1}$ Department of Mechanical and Industrial Engineering, University of Illinois Chicago, Chicago, IL 60607, USA\\
$^{2}$ Department of Ophthalmology, Stritch School of Medicine, Loyola University Chicago, Maywood, IL 60153, USA\\
$^{*}$ These authors contributed equally to this work as co-first authors.\\
$^{\dagger}$ Corresponding author: jiexu@uic.edu
\end{center}

\begin{abstract}
A polarimeter does not report stress. It reports a retardance and an azimuth, and
converting those to a stress pair fixes both the calibration that is required and the
covariance that the subsequent inference must carry. We set out that observation chain for
planar viscoelastic flow, combine it with velocimetry, and ask what the combined
measurement can resolve. The calibration constant is fixed by three separately measurable quantities
(path length, stress-optic coefficient and wavelength) rather than being fitted. Propagating the polarimetric errors to first order gives a stress covariance
that is anisotropic and site dependent even for independent homoscedastic inputs, and
whose conditioning degrades as the retardance approaches zero, where the linearization
itself stops describing the measurement. Wrapping imposes a separate design bound. Two numerical studies follow, both in ideal
calibrated stress coordinates under a prescribed covariance rather than the propagated
polarimetric one. Holding the total scalar count fixed and varying the split between
velocimetric and optical sites, an unequal allocation favoring optical sites outperforms
either pure configuration. We then ask what the measurement resolves between constitutive models compatible
with the same velocity data. In self-consistent
pressure-driven flow of the finitely extensible nonlinear elastic Peterlin model, at
extensibility $L^2=50$ and $3\%$ noise, the optical channel rejects the
velocity-compatible Oldroyd-B family in $78.1\%$ of realizations at the Deborah
number $\mathrm{De}=2$, the ratio of the relaxation time to the flow timescale, and in
$100\%$ at $\mathrm{De}=4$, while rejecting only about $5\%$ below $\mathrm{De}=0.3$.
The resolving power of a design is therefore a strong function of Deborah number and
must be quoted with it.
\end{abstract}


\section{Introduction}\label{sec:intro}
A polarimeter measures light, not stress. Light crossing a birefringent medium
accumulates a phase difference between two polarization components, the retardance, and
emerges with a characteristic orientation, the azimuth. In a flowing polymer solution
that birefringence is produced by the stretching and alignment of the dissolved chains,
so under a stress-optic relation the two readings are set by the local
stress~\cite{maxwell1874,lodge1956,philippoff1961,fuller,janeschitz1983}. The attraction
is practical: nothing is placed in the flow, the measurement is taken through a window,
and it returns a field rather than a point. Polarimetry is used on that basis for
photoelastic stress analysis and polarization
tomography~\cite{hammer2004,lionheart2009,desai2016}, for flow birefringence in polymer
solutions and structured
suspensions~\cite{nakamine2024,worby2024,kawaguchi2025,kobayashi2025}, and in
microfluidic rheo-optics~\cite{quinzani1994,ober2011,salipante2025,haward2013}. The
difficulty is the conversion. A stress-optic relation is material specific, departs from
proportionality as rate and deformation history change~\cite{rothstein2002}, and has
limits that unsteady flow exposes~\cite{noto2025}. Least often stated is what the
conversion does to the measurement uncertainty, and that omission is what this paper sets
out to repair.

Accurate stress measurement matters because so much about a polymer solution
depends on its stress rather than on its motion. Dilute and semi-dilute solutions run
through the processing and handling of complex fluids and through biological fluids;
hyaluronic acid solutions of the kind used to represent synovial fluid are a standard
test case in microfluidic rheometry~\cite{haward2013,meyer2009}. In all of these it is
the stress, not the velocity, that sets the force transmitted to walls and to suspended
particles, that triggers the elastic instabilities limiting throughput, and that a
constitutive model is required to predict~\cite{yamani2023}. Velocity fields are
comparatively easy to obtain and stress fields are not, so the accuracy with which stress
can be recovered, together with the honesty of the uncertainty attached to it, sets the
limit on what any of those questions can be answered with.

A polymer solution is not simply a more viscous liquid. The dissolved chains stretch and
align as the fluid deforms, store elastic energy while stretched, and relax back over a
characteristic time once the deformation stops. While they are stretched they transmit
force, so the fluid carries stress that a Newtonian liquid of the same viscosity does
not. In a planar flow that extra stress is a symmetric two-by-two tensor, and two of its
combinations matter here: the difference between the two in-plane normal components,
written $N_1$ below, and the shear component. A Newtonian liquid in simple shear has no
normal stress difference at all. A polymer solution does, and it is that signature the
optical channel responds to.

How much of it appears depends on how fast the flow deforms the fluid compared with how
fast the chains relax, and that ratio is the Deborah number. Near zero, the chains relax
as quickly as the flow stretches them and the response is almost Newtonian. At order one and above the
stretch persists between deformations, and the elastic contribution becomes
comparable to or larger than the solvent contribution, depending also on the solvent
fraction, the chain extensibility and the flow itself. The Deborah number is
therefore a dial set by the operating point rather than a constant of the fluid, which is
why the resolving power of a measurement has to be quoted at a stated Deborah number
instead of as a single figure.

The relation between deformation and the stress it produces does not follow from the flow
alone. It is supplied by a constitutive model, and several competing models are in routine
use. Two of them can be tuned to reproduce the same velocity profile in the same channel
while predicting different normal stress differences, and a velocimeter cannot tell them
apart, because it sees only the field on which they agree. That is the situation in which
an optical channel stops being a redundant second view of the same information and becomes
the only available criterion. This paper uses that case later to quantify what the combined design can actually
resolve.

The measurement context is a mature one. Calibrated relations that handle those
departures have been established for several material classes and flow
geometries~\cite{nakamine2024,worby2024,kawaguchi2025,kobayashi2025}, and
section~\ref{sec:models} sets out the chain they belong to. Microfluidic rheo-optics and planar-contraction
measurements provide settings where optical and velocity data are acquired
together~\cite{quinzani1994,ober2011,salipante2025,haward2013}, and for the
hyaluronic-acid solutions often used in such devices the stress-optic coefficient has been
determined rheo-optically~\cite{meyer2009}. Finite averaging windows and gaps in processed
correlation maps are part of the forward model rather than
afterthoughts~\cite{haward2013,nekkanti2023}.

The need to interpret motion mechanically also arises in biological flow imaging. Video
microscopy of the conjunctival microcirculation resolves vessel geometry and erythrocyte
motion~\cite{shahidi2010}, related microfluidic measurements characterize erythrocyte
aggregation by cross-correlation and optical flow~\cite{gaynes2022}, and consistent
acquisition establishes a basis for comparing flow metrics across
subjects~\cite{patel2022}. Those studies report kinematics. Converting any such measurement into a mechanical
quantity requires a calibrated forward model and an uncertainty budget of the kind
set out here, though the optical and material model appropriate to whole blood is not
the polymer stress-optic relation used in this paper.

The conversion chain is the object of study here. We set out the conversion and its
error propagation, combine the optical channel with velocimetry, and ask what the combined
measurement can resolve. Four design questions follow. What must be calibrated, and to
what precision. How the polarimetric errors propagate into the stress pair, and where that
propagation becomes ill conditioned. How far the retardance may be driven before fringe
order becomes ambiguous. Finally, for a fixed total number of scalar readings, how those
readings should be divided between the two modalities.

Allocating a fixed measurement budget between modalities is an experimental design
question~\cite{alexanderian,pukelsheim}, and we treat it as one, holding the scalar count
fixed rather than the site count. We then apply the combined measurement to the case set out above, two constitutive
models fitted to the same velocity profile. The resolving power we obtain is a strong
function of Deborah number, which means it cannot be quoted as a single number for the
design.

The paper establishes the observation chain and its consequences for design: the calibration constant and the uncertainty it carries, the stress
covariance the chain induces, and the bound that fringe wrapping imposes. These are
properties of the stated observation model. It then asks what such a measurement buys,
in two numerical studies: how a fixed budget of readings should be split between the
channels, and what the result resolves between constitutive models. Those studies are
run in ideal calibrated stress coordinates under a prescribed covariance; the
polarimetric covariance derived in the first part is not propagated into them.

Section~\ref{sec:models} sets out the observation model, the calibration chain and the
propagated covariance. Section~\ref{sec:allocation} treats allocation between modalities
at fixed scalar count. Section~\ref{sec:channel} demonstrates the resolving power of the
combined measurement on velocity-compatible constitutive models. Derivations and
sensitivity analyses are given in the Supplementary Material.

\section{Observation model and calibration chain}\label{sec:models}

\subsection{Velocity and calibrated stress observations}

Write $\bm\sigma$ for the symmetric polymer extra-stress that the two channels
are being asked to determine. Steady planar creeping flow fixes how it relates to
the velocity, through momentum balance and incompressibility,
\begin{equation}
-\nabla p+\eta_s\Delta\bm u+\nabla\cdot\bm\sigma=\bm0,
\qquad \nabla\cdot\bm u=0.
\label{eq:stokes}
\end{equation}
On an endpoint-free periodic domain at fixed mean velocity this inverts to a
velocity response that is blind to any pressure-like contribution,
\begin{equation}
\mathcal V\bm\sigma=(-\eta_s\Delta)^{-1}
\mathcal P(\nabla\cdot\bm\sigma),
\label{eq:velop}
\end{equation}
with $\mathcal P$ discarding gradient forcing. What particle image velocimetry
(PIV) delivers is not that field itself but its average over an interrogation
window, and the window enters the observation operator explicitly. Once the solvent
contribution and the pressure are specified, the divergence
$\mathcal D\bm\sigma=\nabla\cdot\bm\sigma$ becomes directly observable; recovering
pressure from velocimetry under an assumed Newtonian stress~\cite{vanoudheusden2013}
is one way of supplying that extra information.

The optical channel is described by the calibrated pair
\begin{equation}
\mathcal B\bm\sigma=(b_1,b_2)
=(N_1,2\sigma_{xy}),\qquad N_1=\sigma_{xx}-\sigma_{yy}.
\label{eq:pair}
\end{equation}
which for a spatially resolved field returns the planar deviatoric stress at each
point,
\begin{equation}
\operatorname{dev}\bm\sigma=\mathcal R\bm b
=\frac12\begin{pmatrix}b_1&b_2\\b_2&-b_1\end{pmatrix}.
\label{eq:dense_inverse}
\end{equation}
Equation~\eqref{eq:dense_inverse} serves both as the optical estimator on its own
and as the reference against which the combined measurement is judged. Averaging
both optical components over the same window gives the averaged deviatoric stress,
and each sensor's spatial response appears in the observation matrix rather than
being idealized away. Where this paper says optical data, it means these calibrated
polymer-stress coordinates; the calibration itself is the subject of the next
subsection.

\subsection{Optical calibration and transfer to experiments}

Stress-optic calibration is material-specific. Direct stress and
birefringence measurements exhibit strain- and rate-dependent nonlinearity
and concentration-dependent history effects~\cite{rothstein2002}.
Oscillatory polymer-flow measurements reveal a history-sensitive optical
response relative to local shear rate~\cite{noto2025}. Calibrated
second-order relations resolve optical-axis shear contributions in
cellulose nanocrystal suspensions~\cite{nakamine2024,worby2024,kawaguchi2025},
while quasi-two-dimensional mixed-flow measurements support a calibrated
root-sum-square response to in-plane stress components~\cite{kobayashi2025}.
For the selected finitely extensible nonlinear elastic Peterlin (FENE-P)
convention, in which a chain's extension is capped at a finite maximum, polymer stress
contains the varying factor $f$ multiplying conformation. Consequently, the ideal stress
coordinates used here require an appropriate material calibration before
they represent a raw birefringence experiment.

That calibration chain is worth making explicit, because the stress pair
$\boldsymbol B=(N_1,2\tau_{p,xy})$ used throughout is not what a polarimeter
reports. A polarimeter reports a retardance $\delta$ and an azimuth $\chi$.
Under the stress-optic rule these are related to the in-plane stress pair by
\begin{equation}
 \delta=\kappa\sqrt{N_1^2+(2\tau_{p,xy})^2},\qquad
 \chi=\tfrac12\operatorname{atan2}\!\left(2\tau_{p,xy},\,N_1\right),\qquad
 \kappa=\frac{2\pi d C}{\lambda},
 \label{eq:polarimetric_forward}
\end{equation}
with $d$ the path length, $\lambda$ the wavelength and $C$ the stress-optic
coefficient, so the calibration constant is determined by three separately
measurable quantities rather than a free parameter. Inverting
Eq.~\eqref{eq:polarimetric_forward} gives
$N_1=(\delta/\kappa)\cos2\chi$ and $2\tau_{p,xy}=(\delta/\kappa)\sin2\chi$,
and the measurement covariance of the pair follows from the Jacobian
\begin{equation}
 \boldsymbol J=
 \begin{pmatrix}
  \cos2\chi/\kappa & -2(\delta/\kappa)\sin2\chi\\[2pt]
  \sin2\chi/\kappa & \phantom{-}2(\delta/\kappa)\cos2\chi
 \end{pmatrix},
 \qquad
 \boldsymbol\Sigma_{\boldsymbol B}=\boldsymbol J\,
 \boldsymbol\Sigma_{(\delta,\chi)}\,\boldsymbol J^T .
 \label{eq:polarimetric_covariance}
\end{equation}

\begin{figure}[tbp]
\centering
\includegraphics[width=0.72\textwidth]{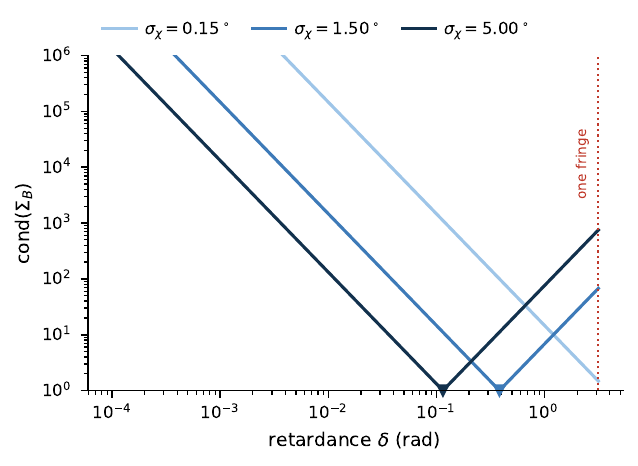}
\caption{\label{fig:covariance}
Conditioning of the propagated stress covariance $\boldsymbol\Sigma_{\boldsymbol B}$ as a function of retardance. The Jacobian of Eq.~\eqref{eq:polarimetric_covariance} has orthogonal columns, one radial and one tangential in the $(N_1,2\tau_{p,xy})$ plane, so the eigenvalues are $\sigma_\delta^2/\kappa^2$ and $4\delta^2\sigma_\chi^2/\kappa^2$ exactly, and the condition number is independent of $\kappa$ and of the azimuth. It diverges as $\delta\to0$, where the first-order description itself stops applying: the linearization omits the product of retardance and azimuth noise, so the vanishing eigenvalue is a limit of validity rather than a physical statement about a low-stress site. Markers give the crossover $\delta^{*}=\sigma_\delta/2\sigma_\chi$ at which the two eigenvalues coincide, shown for $\sigma_\delta=0.02$ rad and the three labeled azimuth noise levels. The dotted line is the single-fringe bound.
}
\end{figure}

Two consequences are worth recording (Figure~\ref{fig:covariance}). First,
$\boldsymbol\Sigma_{\boldsymbol B}$ is anisotropic and site dependent even when
$(\delta,\chi)$ errors are independent and homoscedastic, and its conditioning
degrades as $\delta\to0$. The vanishing eigenvalue is a property of the
linearization rather than of the instrument: it reports zero uncertainty along the
tangential direction, which inverse-covariance weighting would then weight heavily.
The first-order form omits the product of retardance and azimuth noise. For
independent Gaussian errors in a locally signed $\delta$ the exact tangential
variance is
$(\delta^2+\sigma_\delta^2)\left[1-e^{-8\sigma_\chi^2}\right]/(2\kappa^2)$.
Their ratio is
$[1+(\sigma_\delta/\delta)^2]\left[1-e^{-8\sigma_\chi^2}\right]/(8\sigma_\chi^2)$,
so the first-order expression underestimates the variance by a factor of $1.99$ at
$\delta=\sigma_\delta$, rising to $100.5$ at $\delta=0.1\,\sigma_\delta$ and
$1.1\times10^{3}$ at $\delta=0.03\,\sigma_\delta$, quoted here at
$\sigma_\chi=2^\circ$. It is therefore only at retardances well below the
retardance noise that the error reaches orders of magnitude. A design that places sites at low retardance
therefore needs the underlying intensity or Stokes likelihood rather than this
covariance, which is the limit of validity rather than a physical loss of information. Second, the azimuth is defined modulo $\pi$, and the retardance is recovered on a
principal branch whose extent depends on the polarimeter. Take that branch as
$\delta<\pi$, which folds the fast- and slow-axis sign ambiguity into the azimuth.
A design must then either respect the bound, equivalently
$dC\,\lvert\boldsymbol B\rvert<\lambda/2$, or supply independent fringe-order
information. The instrument's own branch and unwrapping convention set the bound in
any particular case. For hyaluronic acid in buffered
saline, $C=1.82\times10^{-8}\,$Pa$^{-1}$ has been determined rheo-optically
and reported as independent of concentration, molar mass and shear rate~\cite{meyer2009}; it is the value adopted in the
cross-slot measurements of Ref.~\cite{haward2013}. With $\lambda=546$ nm and
$d=2.1$ mm this gives $\kappa\approx4.4\times10^{-4}\,$rad\,Pa$^{-1}$, so
the single-fringe condition corresponds to in-plane stress differences below
several kPa. Stating the chain and its covariance in this form is what connects the planar
operator used throughout to a polarimetric observable without ambiguity, and it
fixes which quantities a calibration would have to supply.

Depth averaging and optical-axis coverage are also part of the forward
model. A known, nonzero gain multiplying a separable planar field preserves
its operator kernel. Nonseparable depth dependence and three-dimensional
ray geometry require the corresponding tensor-ray
description~\cite{sharafutdinov1994,hammer2004,lionheart2009,desai2016}.
The Supplementary Material records the planar depth-gain argument and the
scope of three-dimensional rank counting. Those structural checks specify
the geometric conditions for transferring the planar observation model
to an optical experiment.

The numerical studies use prescribed synthetic covariance and calibrated
stress coordinates. Their error intervals and compatibility tests are
frequentist simulation summaries under those settings. Transfer to an
instrument uses measured joint covariance, spatial response and calibration
uncertainty in the same observation model.

\section{Measurement allocation}\label{sec:allocation}

\subsection{Equal-count allocation between modalities}

The allocation study reuses the reconstruction machinery
of~\cite{liu2026identifiability}, with the settings fixed as follows. Each of
$40$ truth fields is a periodic deviatoric tensor field with component cutoff $K=8$,
spectral length $0.24$, spectral power $3$ and full-tensor Airy energy fraction
$0.35$, evaluated at the physical sites without interpolation. Eighty-one
two-coordinate site assignments give $162$ scalar readings. The sites are the leading
entries of one deterministic periodic farthest-point ordering of a $17\times17$
candidate grid, and each modality takes a prefix of that same ordering, so the five
splits differ in allocation rather than in placement family. The fixed-box branch
averages the truth over centered rectangular apertures of width $8/17$ for
velocimetry and $2/17$ for optics; the point branch samples the same truths
pointwise. Noise is spatially independent between sites, and the arms of a pair share their
realized noise. Its scale is set once, from a separately generated fixed reference
field rather than from the truth ensemble: that field fixes a pooled velocity scale
$s_v=0.0483$ and a pooled optical scale $s_o=0.3941$, and the per-site covariances are
\begin{equation}
\boldsymbol\Sigma_v=(0.03\,s_v)^2\begin{pmatrix}1&0.20\\0.20&1\end{pmatrix},
\qquad
\boldsymbol\Sigma_{\boldsymbol B}
=(0.03\,s_o)^2\begin{pmatrix}1&-0.50\\-0.50&4\end{pmatrix},
\label{eq:allocation_covariance}
\end{equation}
held fixed across allocations and across the point and box branches. This is the
declared covariance that whitens the problem, and it is not the independent
homoscedastic one of the channel example: the within-site correlations are $+0.20$ and
$-0.25$, and the second optical coordinate carries twice the standard deviation of the
first. Each reconstruction minimizes
$\lVert B\bm w-W\bm y\rVert_2^2+\alpha\lVert\bm w\rVert_2^2$ with $B=W\mathcal O R_\ell$.
Here $W$ is the Cholesky whitener of the declared covariance and $R_\ell$ the inverse
factor of the periodic $H^1$ penalty at length $\ell=0.15$. The parameter $\alpha$
runs over $121$ logarithmic values from $10^{-14}$ to $10^{6}$ times
$\sigma_{\max}(B)^2$. The reported error is the relative Frobenius error of the
deviatoric part, which removes the isotropic gauge.

\begin{figure}[tbp]
\centering
\includegraphics[width=0.72\textwidth]{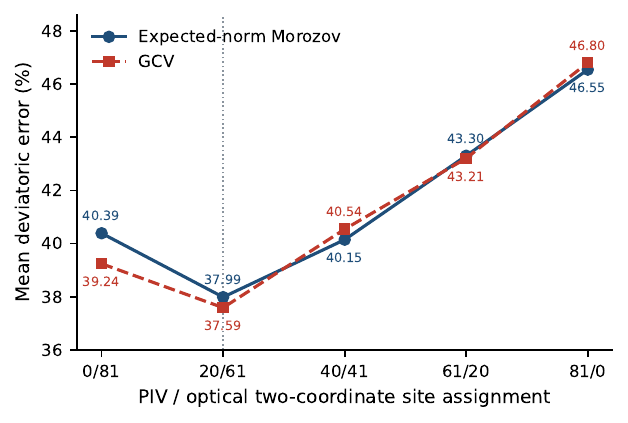}
\caption{\label{fig:allocation}
Equal-count allocation between velocimetric and optical sites, from supplement table S6 at reconstruction grid $n=65$. Labels give the PIV/optical two-coordinate site split; every configuration carries $162$ scalar readings at fixed physical apertures. The dotted line marks the allocation of lowest mean error, which is the same under both rules. Among the five splits tested the lowest-error allocation is an interior one under both rules, so at fixed measurement count neither pure configuration is the best of those tried. The five are a comparison at fixed placement family, not a continuous optimization.
}
\end{figure}

Figure~\ref{fig:allocation} gives the comparison. Both reported rules choose the regularization parameter from the data.
Expected-norm Morozov matches the whitened residual norm to the value expected from
the declared noise. Generalized cross-validation (GCV) minimizes a rotation-invariant approximation to the
leave-one-out prediction score and needs no declared noise amplitude, although the
relative covariance that whitens the problem enters both rules alike.
Holding the scalar count fixed at $162$ readings and varying only the split
between modalities, the best allocation is neither pure configuration. An unequal split
favoring optical sites outperforms all-optical by $2.403$ percentage points under Morozov,
with pointwise $95\%$ interval $[2.068,2.738]$, and by $1.654$ percentage points under GCV,
interval $[1.097,2.212]$. The ordering is the same for both rules and at every
reconstruction cutoff tested. Two qualifications belong with the number. The intervals are
pointwise and unadjusted for comparing five allocations, and equal scalar count is a
controlled comparison rather than a cost model: instrument time, exposure and processing
each define further design resources~\cite{alexanderian,pukelsheim}.

\subsection{Measurement allocation and model assessment}

The observation structure explains the complementary role of velocity in
reconstructing incomplete optical stress maps. Optical measurements supply
local deviatoric components, while momentum balance couples their spatial
variations. In the finite-aperture stripe experiment, adding velocity
reduces whole-domain error from $50.40\%$ to $27.82\%$~\cite{liu2026identifiability} under the stated
synthetic covariance. The shared-gap and fixed-aperture refinement results
support this interpretation. With complete optical coverage, the GCV
comparison gives essentially unchanged error; the contribution of velocity
depends on measurement coverage and regularization. The separate equal-count
experiment isolates a sampling-allocation effect within one placement
family. Actual acquisition resources can be represented through modality-
specific time, exposure, processing and precision models~\cite{alexanderian,pukelsheim}.

Normal stress provides a complementary criterion for constitutive-model
assessment. The candidate family throughout is Oldroyd-B, the simplest
viscoelastic description of a dilute solution: a Newtonian solvent carrying one
elastic mode with a single relaxation time and no bound on how far a chain can
stretch. In steady Oldroyd-B channel flow, relaxation time changes the
normal stress while velocity depends on the viscosities. The FENE-P channel
study exploits this distinction: velocity-compatible Oldroyd-B fits retain
a measurable normal-stress shape discrepancy at higher Deborah number.
Profiling relaxation time against optical data tests compatibility of the
candidate family, with the parameter sweep and solvent contribution
determining the diagnostic regime.

These examples connect the operator-level distinction between measured and
modeled stress to two practical decisions. The reconstruction experiment evaluates
how velocity supports incomplete optical stress maps. The channel experiment evaluates
whether a constitutive model consistent with kinematics also accounts for
stress. Their common framework is the physical observation model, while
their domain, stress ensemble and noise calibration remain explicit.

\subsection{Domains and decision criteria}

The reconstruction study uses an endpoint-free torus, fixed physical
apertures and uniform physical mass. The model-assessment study uses a bounded no-slip channel with a known
pressure gradient and solvent viscosity. Each experiment specifies its
observation operator, covariance and decision criterion. Finite-window sampling and explicit boundary information are treated in the
Supplementary Material as observation settings in their own right, with
trapezoidal window mass and pressure elimination referred to the full discrete
gradient range.

The reconstruction comparison fixes observation operators, methods and
paired noise draws before evaluating performance. The channel comparison fixes the
constitutive family, nuisance-parameter treatment and null-test threshold
before scanning the parameter grid. The resulting error differences and
model-compatibility decisions quantify distinct uses of stress data.

\paragraph{Numerical implementation}
Ranks are counted as the number of singular values exceeding
$\tau=\max(m,n)\epsilon_{\rm mach}\sigma_{\max}$ for an $m\times n$ matrix.
That threshold serves only as a rank diagnostic; the regularized reconstruction
itself keeps every singular value. Sensitivity to a relative threshold is reported
separately as a conditioning check. The Supplementary Material holds the
absolute-threshold and finite-field rank witnesses, together with the environment,
seeds, covariance, regularization choices and residual checks behind every figure.

\section{Normal-stress diagnostics in pressure-driven channel flow}\label{sec:channel}

Normal-stress profiles provide a constitutive diagnostic when velocity
measurements admit several candidate models. We quantify this additional
information in fully developed channel flow with known pressure gradient
and solvent viscosity. Classical analytical solutions~\cite{cruz2005}
provide self-consistent velocity and stress fields, allowing the normal-
and shear-stress contributions to model discrimination to be evaluated
separately.

\subsection{Self-consistent flow and observation model}

Consider walls at $y=\pm H$, pressure $p=-Gx$, and velocity
$\boldsymbol u=(u(y),0,0)$ with no slip. The equilibrium-normalized, three-dimensional FENE-P model, in its
nondiffusive form, is~\cite{yamani2023}
\begin{equation}
\begin{split}
 D_t\boldsymbol A-(\nabla\boldsymbol u)\boldsymbol A
 -\boldsymbol A(\nabla\boldsymbol u)^T
 &=-\lambda^{-1}(f\boldsymbol A-\boldsymbol I),\\
 f=\frac{L^2-3}{L^2-\operatorname{tr}\boldsymbol A},\qquad
 \boldsymbol\tau_p=\boldsymbol\sigma
 &=\frac{\eta_p}{\lambda}(f\boldsymbol A-\boldsymbol I).
\end{split}
\label{eq:channel_fene}
\end{equation}
For signed shear $g=u'(y)$, steady balance
gives $A_{yy}=A_{zz}=1/f$, $A_{xy}=\lambda g/f^2$,
$A_{xx}=1/f+2(\lambda g)^2/f^3$, and $A_{xz}=A_{yz}=0$.
Thus the out-of-plane component satisfies the same constitutive dynamics as
the in-plane components. The closure and momentum equations reduce to
\begin{equation}
\begin{split}
 f^2(f-1)&=\frac{2(\lambda g)^2}{L^2},\qquad
 (\eta_s+\eta_p/f)g=-Gy,\\
 \tau_{p,xy}&=\eta_p g/f,\qquad
 N_1=\tau_{p,xx}-\tau_{p,yy}=2\eta_p\lambda g^2/f^2.
\end{split}
\label{eq:channel_balance}
\end{equation}
Setting $r=|g|/f$ yields the monotone cubic
$(\eta_s+\eta_p)r+(2\eta_s\lambda^2/L^2)r^3=G|y|$.
We solve its unique nonnegative root and integrate $|g|$ from $|y|$ to $H$
to obtain velocity. An independent root calculation in $f$ and a closed-form
velocity primitive verify the solution.

Lengths, stresses, and velocities are scaled by $H$, $GH$, and
$U_0=GH^2/\eta_0$, where $\eta_0=\eta_s+\eta_p$.
These reference scales are held fixed when fitting alternative constitutive parameters.
The primary set uses $G=H=\eta_0=1$, $\eta_s=\eta_p=0.5$,
$L^2=50$, and nominal $\mathrm{De}=\lambda GH/\eta_0$
equal to $0.1$, $0.3$, $1$, $2$, and $4$.
Table~\ref{tab:channel} also reports the actual wall value
$\mathrm{Wi}_{\rm wall}=\lambda\max_y|g|$.
The full parameter map includes $L^2=25,100$ and noise levels of $1\%,3\%,5\%$.
The Supplementary Material tabulates velocity non-rejection counts and both
unconditional and conditional optical-family rejections across this map.

\begin{figure}[tbp]
\centering
\includegraphics[width=0.72\textwidth]{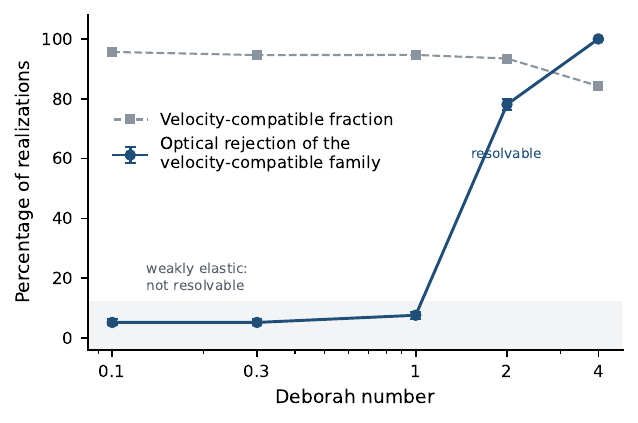}
\caption{\label{fig:de_trend}
Constitutive resolving power of the combined measurement against Deborah number, from Table~\ref{tab:channel}. Circles give the percentage of velocity-compatible realizations in which the optical channel rejects the fitted Oldroyd-B family, with pointwise $95\%$ Wilson intervals; squares give the fraction of realizations the velocity test retains. Each point uses 2\,000 realizations at $L^2=50$ and $3\%$ noise.
}
\end{figure}

The strongest trend in that map is in the Deborah number, and it is reported
here rather than left to the Supplementary Material because it determines how the design's
resolving power may be quoted. Figure~\ref{fig:de_trend} and Table~\ref{tab:channel} give
it: at $L^2=50$ and $3\%$ noise the conditional rejection rate rises from near the
significance floor below $\mathrm{De}=0.3$ to $78.1\%$ at $\mathrm{De}=2$ and $100\%$ at
$\mathrm{De}=4$. Resolving power is therefore not a fixed property of the observation
design, and any statement of it must name the Deborah number it refers to.

We observe streamwise velocity and the calibrated polymer-stress pair
$\boldsymbol B=(N_1,2\tau_{p,xy})$ at 31 equally spaced positions
$y/H\in[-0.9,0.9]$. Independent Gaussian noise is added across sites,
coordinates, and modalities. The velocity standard deviation is the specified
fraction of the true velocity RMS. Both stress coordinates share a standard
deviation given by that fraction of
$[\sum_i(N_{1,i}^2+4\tau_{p,xy,i}^2)/(2n)]^{1/2}$, with $n=31$.
We denote these standard deviations by $s_u$ and $s_B$, respectively.
These are ideal, calibrated stress observations. Finite extensibility introduces
the spatially varying factor $f$ between conformation anisotropy and stress;
experimental birefringence therefore requires material-specific calibration.

\subsection{Velocity fitting and stress-family compatibility}

The candidate Oldroyd-B model fits positive polymer viscosity to velocity,
with $G$ and $\eta_s$ fixed. In the dimensionless channel the streamwise
profile is $u_i=a\,v_i$, with fixed shape $v_i=(1-y_i^2)/2$ and a single
amplitude $a$ that carries the total viscosity,
\begin{equation}
 a=\eta_0^{-1},\qquad \eta_0=\eta_s+\eta_p .
 \label{eq:amplitude_viscosity}
\end{equation}
The reference scaling above sets $\eta_0=1$ for the primary parameter set, so
the \emph{nominal} amplitude there is $a=1$; this is a consequence of the
normalization and not an independent result. We nonetheless estimate the
amplitude from the velocity data rather than imposing it, because only the
estimate carries the uncertainty that the stress-family test consumes. The
least-squares estimate and its variance are
$\widehat a=\boldsymbol v^T\boldsymbol u^{\rm obs}/
(\boldsymbol v^T\boldsymbol v)$ and
$s_a^2=s_u^2/(\boldsymbol v^T\boldsymbol v)$;
equivalently the fit returns $\widehat\eta_0=\widehat a^{-1}$ and hence
$\widehat\eta_p=\widehat\eta_0-\eta_s$. Every expression below uses
$\widehat a$, never the nominal value, so no step of the test presumes
$a=1$.

Because the velocity noise level is specified as a fraction $c$ of the true
velocity root-mean-square, the amplitude standard error simplifies. Writing
$s_u=c\,a\,\lVert\boldsymbol v\rVert/\sqrt n$ gives
\begin{equation}
 \frac{s_a}{a}=\frac{c}{\sqrt n},
 \label{eq:amplitude_relative_error}
\end{equation}
for a matched profile. With $n=31$ sites this is $0.180\%$, $0.539\%$ and
$0.898\%$ at the $1\%$, $3\%$ and $5\%$ noise levels respectively: averaging
over the sites reduces the random amplitude error by $\sqrt n\approx5.6$ relative to
the noise on a single site. That reduction does not make the amplitude negligible,
because its error is shared by every site. The propagated rank-one term
$s_a^2\boldsymbol h\boldsymbol h^T$ in Eq.~\eqref{eq:channel_covariance} has one
nonzero eigenvalue. Relative to $s_{\boldsymbol B}^2$ it equals $1.99$, $1.92$,
$1.39$, $0.81$ and $0.44$ at $\mathrm{De}=0.1$, $0.3$, $1$, $2$ and $4$, so it
raises the variance along $\boldsymbol h$ by $44\%$ to $199\%$. It is retained for that
reason. Equation~\eqref{eq:amplitude_relative_error} is exact when the true profile
is $a\boldsymbol v$; under the finitely extensible alternative the relative error
refers to the projected amplitude, a correction below $0.0001$ percentage points at
$\mathrm{De}=4$.
This velocity fit identifies viscosity while leaving relaxation time
$\lambda$ free: all positive values have the same velocity likelihood.

We therefore test compatibility with the entire positive-$\lambda$ Oldroyd-B
family. Its normal stress is $N_{1,i}=c y_i^2$, with
$c=2\eta_p\lambda\widehat a^2\geq0$ fitted to the stress observations.
Its shear prediction is $2\tau_{p,xy,i}=-2y_i+2\eta_s y_i\widehat a$.
Propagation of velocity-fit uncertainty gives the shear residual covariance
\begin{equation}
 \boldsymbol C_s=s_B^2\boldsymbol I+s_a^2\boldsymbol h\boldsymbol h^T,
 \qquad h_i=2\eta_s y_i.
 \label{eq:channel_covariance}
\end{equation}
The optical-family score is the sum of the squared whitened shear and
profiled normal-stress residual norms. Stress data determine the nuisance
normal-stress amplitude, so the score assesses compatibility with the
constitutive family after profiling relaxation time against the stress data.

For interior fits, the velocity, normal-shape, shear, and combined scores have
$30$, $30$, $31$, and $61$ Gaussian residual degrees of freedom, respectively.
Before the FENE-P sweep, 2\,000 matched-null simulations repeated the velocity
and stress nuisance fits. Their empirical 95th-percentile critical values are
$44.171$, $44.351$, $45.296$, and $80.438$ in the same order.
All calibration and evaluation fits remained inside the positivity bounds.
The evaluation uses 2\,000 independent noise realizations at each of the
45 FENE-P settings, five matched Oldroyd-B controls and three zero-solvent
controls. Case-specific streams are independent of each other and of the
null calibration. The fixed empirical thresholds have a nominal pointwise
$5\%$ level, and rate intervals use the Wilson $95\%$ construction.
The Supplementary Material records the independent validation design,
analytical powers and the separate 40-realization exploratory map.
A separate 2\,000-replicate Newtonian null calibrates the zero-$N_1$ comparison.

\subsection{Normal-stress discrimination and parameter window}

At $3\%$ noise and $L^2=50$ the combined measurement identifies a clear diagnostic
window at $\mathrm{De}=2$ and $4$ (Table~\ref{tab:channel}). The two-stage structure is
what matters for design. The velocity test retains most realizations throughout, so it
does not by itself separate the two models; the optical channel then rejects the fitted
family within that retained set, and only above the elasticity threshold. Velocity establishes
admissibility and optics supplies the discrimination, and the two roles should be read
separately. A free-optics comparator that fits its own shear amplitude without any
velocity measurement rejects $1563/2000$ realizations at $\mathrm{De}=2$ and all
$2000$ at $\mathrm{De}=4$, essentially the conditional rates of the combined test
(supplement table S3). The channel study therefore shows that optical stress
data supply a constitutive criterion that velocity alone lacks; it does not show that
the combined instrument is required for this discrimination. What velocity contributes
here is the admissibility constraint and the viscosity, and its value to the combined
design is established by the reconstruction allocation rather than by this test.

\begin{table}[tbp]
\centering
\caption{Complete primary channel results at $L^2=50$ and $3\%$ noise.
Velocity lists non-rejections and Optics lists rejections of the profiled
Oldroyd-B family, each out of 2\,000 independent realizations per case.
Figure~\ref{fig:de_trend} plots the rejection percentage among velocity
non-rejections that these counts give, with pointwise Wilson $95\%$ intervals.
Tests use fixed empirical thresholds calibrated at the nominal $5\%$ level.}
\label{tab:channel}
\small
\setlength{\tabcolsep}{5pt}
\begin{tabular}{rrrrl}
\toprule
$\mathrm{De}$ & $\mathrm{Wi}_{\rm wall}$ & Velocity & Optics & Conditional optics (\%) \\
\midrule
0.1 & 0.100 & 1913/2000 & 104/2000 & 5.2 [4.3, 6.3] \\
0.3 & 0.301 & 1892/2000 & 107/2000 & 5.2 [4.3, 6.3] \\
1 & 1.019 & 1893/2000 & 149/2000 & 7.6 [6.4, 8.8] \\
2 & 2.131 & 1869/2000 & 1560/2000 & 78.1 [76.2, 79.9] \\
4 & 4.711 & 1685/2000 & 2000/2000 & 100.0 [99.8, 100.0] \\
\bottomrule
\end{tabular}

\end{table}

\begin{figure}[tbp]
\centering
\includegraphics[width=\textwidth]{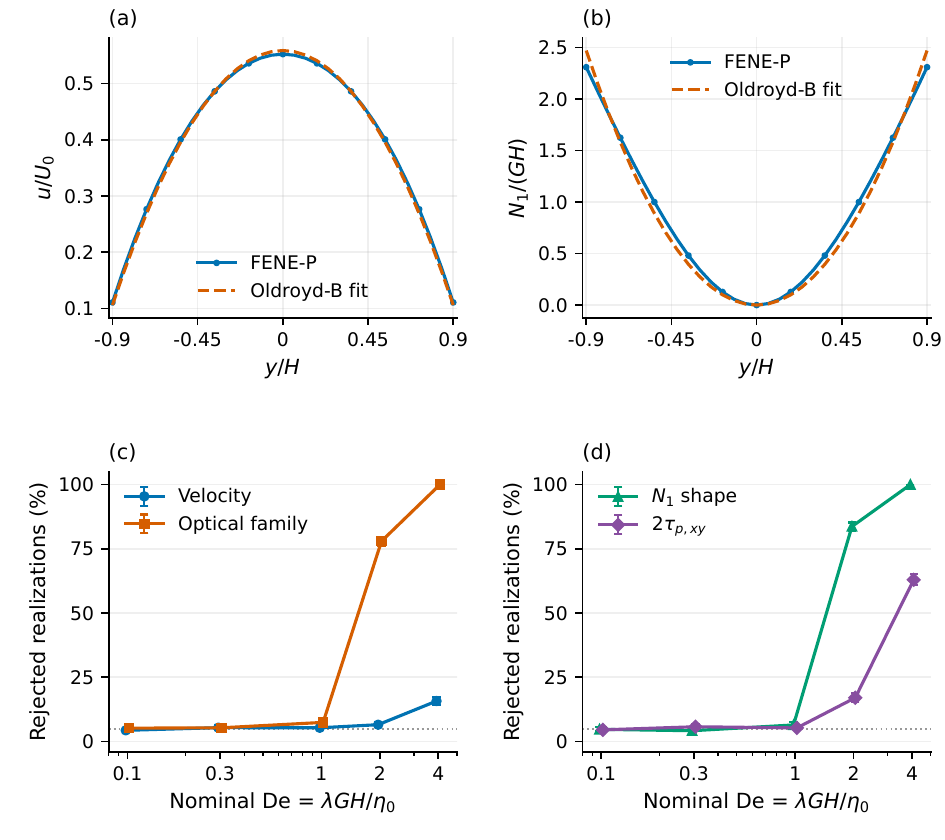}
\caption{Normal-stress diagnosis of constitutive response at $L^2=50$.
(a,b) Illustrative $\mathrm{De}=4$ profiles at the observation sites:
Oldroyd-B viscosity minimizes the noiseless velocity error, and relaxation
time minimizes the normal-stress error. (c) Velocity and optical-family
rejection rates over the complete $3\%$-noise primary sweep.
(d) Separate normal-shape and shear tests identify the source of stress
discrepancy. Error bars are pointwise Wilson $95\%$ intervals from 2\,000
independent realizations per case; dotted lines mark the nominal $5\%$ level.}
\label{fig:channel}
\end{figure}

Normal-stress shape is the stronger of the two optical components, rejecting
$83.70\%$ of realizations at $\mathrm{De}=2$ and $100\%$ at $\mathrm{De}=4$
against $17.05\%$ and $62.95\%$ for the shear component, both unconditional over
$2\,000$ realizations at their own thresholds. That component carries the
finite-extensibility signature, and the gap between the two is the reason an optical
measurement adds a criterion that velocity cannot: a shear-only optical reading would
be substantially weaker at $\mathrm{De}=2$. Figure~\ref{fig:channel} shows the underlying profiles, the
rejection rates over the complete sweep, and the separate normal-shape and shear
tests.

The Gaussian residual laws provide an analytical check of these rates.
The fitted velocity amplitude and the velocity-shape residual occupy
orthogonal Gaussian directions. Thus, for interior fits, conditioning on
velocity non-rejection preserves the population optical rejection
probability. At $\mathrm{De}=2$, the analytical probability at the fixed
empirical optical threshold is $78.73\%$, consistent with the independent
validation. Replacing the empirical cutoffs by analytical $95$th-percentile
cutoffs gives $78.46\%$ in the same validation sample. The Supplementary
Material derives the residual laws, bounds the effect of positivity
constraints, and gives the full extensibility and noise map with explicit
conditional denominators.

Matched-model and zero-solvent controls clarify the physical basis of this
diagnostic. With fixed Oldroyd-B viscosities and pressure, varying
$\lambda$ gives identical velocity and shear but different $N_1$:
these are exact constitutively admissible velocity-indistinguishable states.
Across the five matched Oldroyd-B controls at $3\%$ noise, independent
velocity rejection rates are $3.70$--$5.20\%$ and optical-family rates
are $4.35$--$5.50\%$. With $\eta_s=0$, both constitutive laws satisfy
$\tau_{p,xy}=-Gy$ and $N_1=2(\lambda/\eta_p)\tau_{p,xy}^2$.
The fitted viscosity can then be accompanied by a relaxation time that
matches both optical fields exactly. The zero-solvent check at
$L^2=50$, $\mathrm{De}=1$, $\eta_p=1$ and $3\%$ noise gives
110 optical-family rejections out of 2\,000 ($5.50\%$).

Verification used 65--513 grid points and Gauss--Legendre orders 8--64.
The maximum scaled full-$3\times3$ constitutive residual was
$2.06\times10^{-15}$, momentum imbalance $3.34\times10^{-16}$, and
order-64 velocity quadrature relative error $4.72\times10^{-16}$.
All conformation tensors were positive definite with
$\operatorname{tr}\boldsymbol A<L^2$.
Normal-stress shape supplies the principal additional diagnostic in the
velocity-compatible regimes identified by the channel study.

\section{Discussion}\label{sec:discussion}
Three points are worth separating before the conclusions, because they constrain how the
numbers above should be read.

The first concerns the calibration constant. Writing $\kappa=2\pi dC/\lambda$ in terms
of path length, stress-optic coefficient and wavelength makes it measurable rather than
fitted, and it localizes the uncertainty in three quantities that can be budgeted
separately. For the hyaluronic-acid system the coefficient has been reported
rheo-optically as concentration, molar mass and shear rate independent, and it is reported
without an uncertainty~\cite{meyer2009}; path length and wavelength can be measured, but
how far they contribute is a question for a specific budget rather than something to
assume away. Because $\kappa$ enters as a common gain on the whole stress field, any
design whose criterion depends on the absolute stress scale inherits that uncertainty.
A comparison constructed to be invariant to a common scale escapes it, and the
normal-shape statistic reported here is one: it is normalized by an optical scale that
moves with the data, so a common gain cancels. The combined test used here includes a
predictive-shear component and is therefore not invariant to that gain.

The second concerns where the propagated covariance fails. The singularity at vanishing
retardance is a property of the linearization, not of the instrument: the first-order form
drops the product of retardance and azimuth noise, and the tangential variance it reports
goes to zero while the exact variance does not. What follows is a caution about the
covariance rather than a claim about the measurement. Inverse-covariance weighting would
give such a site a large weight on the strength of a variance that is an artifact, so a
design placing many sites in quiescent regions should not be evaluated with this
covariance. Whether those sites carry usable information is a question for the intensity
or Stokes likelihood, which retains the terms the linearization drops.

The third concerns the scope of the resolving-power result. It is obtained for one
constitutive pair, one flow, and one noise level, and it is a strong function of Deborah
number. It should be read as establishing that the optical channel supplies a criterion
that velocity alone cannot, and as locating the regime in which that criterion is
informative, rather than as a general statement about constitutive discrimination.

\section{Conclusions}

Converting a polarimetric reading into a stress pair is the measurement, not a
preliminary to it. Writing the conversion explicitly fixes what must be calibrated and
what uncertainty the downstream inference is obliged to carry, and both consequences are
quantitative.

The calibration constant is $\kappa=2\pi dC/\lambda$, set by path length, stress-optic
coefficient and wavelength. It is therefore measurable rather than fitted, and its
uncertainty decomposes into three budgetable contributions. For hyaluronic acid in
buffered saline the coefficient has been determined rheo-optically and reported as
independent of concentration, molar mass and shear rate, but without an uncertainty, so a
design that depends on the absolute stress scale inherits an uncertainty the literature
does not quantify.

Propagating the polarimetric errors gives a stress covariance that is anisotropic and site
dependent even when the retardance and azimuth errors are independent and homoscedastic.
Because the Jacobian has orthogonal radial and tangential columns, its eigenvalues are
available in closed form, the condition number is independent of the calibration constant
and of the azimuth, and it diverges as the retardance vanishes. The divergence is a limit of the first-order description rather than a statement about
the instrument, since the linearization omits the product of retardance and azimuth noise.
A design that places many sites at low retardance therefore cannot be evaluated with this
covariance, and needs the intensity or Stokes likelihood instead.
Wrapping imposes a separate and independent bound, since the azimuth is defined modulo
$\pi$ and the retardance modulo one fringe.

Holding the total number of scalar readings fixed rather than the number of sites, an
unequal allocation favoring optical over velocimetric sites outperforms either pure
configuration. Equal scalar count is a controlled comparison, not a cost model.

Applied to constitutive models fitted to the same velocity profile, the combined
measurement supplies a criterion that velocity alone cannot. Its resolving power is
strongly dependent on Deborah number, rising from near the significance floor below
$\mathrm{De}=0.3$ to essentially certain rejection by $\mathrm{De}=4$. The design's
resolving power is accordingly a property of the operating point rather than of the
instrument, and it cannot be reported without the Deborah number it was measured at.

The covariance structure, the wrapping bound and the allocation comparison follow
from the stated observation model and are the parts intended to transfer; the resolving-power figures
are properties of the particular constitutive pair, flow and noise level used to
demonstrate them.

\roles{%
\textbf{Zijian Liu:} Conceptualization, Methodology, Software, Formal
analysis, Investigation, Funding acquisition, Writing -- original draft,
Writing -- review and editing.
\textbf{Julian Olszewski:} Conceptualization, Methodology, Software, Formal
analysis, Investigation, Funding acquisition, Writing -- original draft,
Writing -- review and editing.
\textbf{Bruce I. Gaynes:} Conceptualization, Writing -- review and editing.
\textbf{Jie Xu:} Conceptualization, Methodology, Formal analysis, Investigation, Supervision,
Writing -- review and editing.%
}

\ack{%
The authors declare no conflicts of interest.
The authors used GPT-6 (OpenAI) and Claude Opus 5 (Anthropic) to assist with
literature search, coding, language, organization, and readability. The
authors reviewed and edited the resulting material as needed and take full
responsibility for the content of the publication.%
}

\funding{%
Z.L.\ acknowledges support from the Illinois Society for the Prevention of Blindness (ISPB).
J.O.\ acknowledges support from the University of Illinois Chicago
Chancellor's Undergraduate Research Award (CURA).
B.I.G.\ acknowledges support from the Richard A. Perritt, MD Charitable
Foundation.%
}

\data{%
All data that support the findings of this study are included within the article
and its supplementary information files.%
}

\begingroup\footnotesize\setlength{\bibsep}{0pt}\raggedright\interlinepenalty=10000

\endgroup
\end{document}